\documentclass[aps,pre,twocolumn,superscriptaddress,floatfix,nofootinbib]{revtex4-2}

\usepackage{amsmath,amssymb}
\usepackage{graphicx}
\usepackage{dcolumn}
\usepackage{siunitx}
\usepackage{bm}
\usepackage[colorlinks=true,allcolors=blue]{hyperref}

\newcolumntype{d}{D{.}{.}{-1}}  
\newcommand{\Bt}{\mathrm{B}}
\newcommand{\Om}{\Omega}

\begin{document}

\title{Community structure of the pseudofractal web}

\author{Alexei Vazquez}
\email{alexei.vazquez@gmail.com}
\affiliation{Nodes \& Links Ltd, Salisbury House, Station Road,
Cambridge, CB1 2LA, UK}

\date{\today}

\begin{abstract}
The Ramsey community number $r_\kappa$ is the smallest network size at which a
graph is better described by a partition into communities than by no partition,
under a prescribed detection rule. On a scale-free graph this question is
confounded: a block model can split the network merely to absorb its degree
distribution. I compute $r_\kappa$ \emph{analytically} for the deterministic
pseudofractal scale-free web of Dorogovtsev, Goltsev, and Mendes, separating
genuine community structure from degree heterogeneity with two closed-form
detection rules. Under a plain Bernoulli stochastic block model, the web's
natural recursive bipartition is unpreferred while small and breaks at
$r_\kappa=1095$ nodes, with a log-evidence growing as
$(\ln 3-\tfrac{2}{3}\ln 2)n$. Under a degree-corrected model tested against the
configuration-model null, the same partition survives, breaking far earlier at
$r_\kappa=42$, with a log-evidence growing as $(2\ln 3-\tfrac{4}{3}\ln 2)n$---
exactly twice the plain slope, and independent of the prior. Degree correction
reverses the ordering of the candidate cuts, demoting the hub--leaf split and
elevating the recursive one. Because the web is self-similar, the best
description is not two communities but a nested hierarchy: the degree-corrected
evidence keeps rising as the partition is refined, and is maximised at of order
$\sqrt{n}$ communities of $\sim\sqrt{n}$ nodes. A purely local recursive rule
thus builds true hierarchical community structure, over and above the
scale-free degree sequence it also produces, in an exactly solvable setting.
\end{abstract}

\maketitle

\begin{figure*}[t]
\centering
\includegraphics[width=\textwidth]{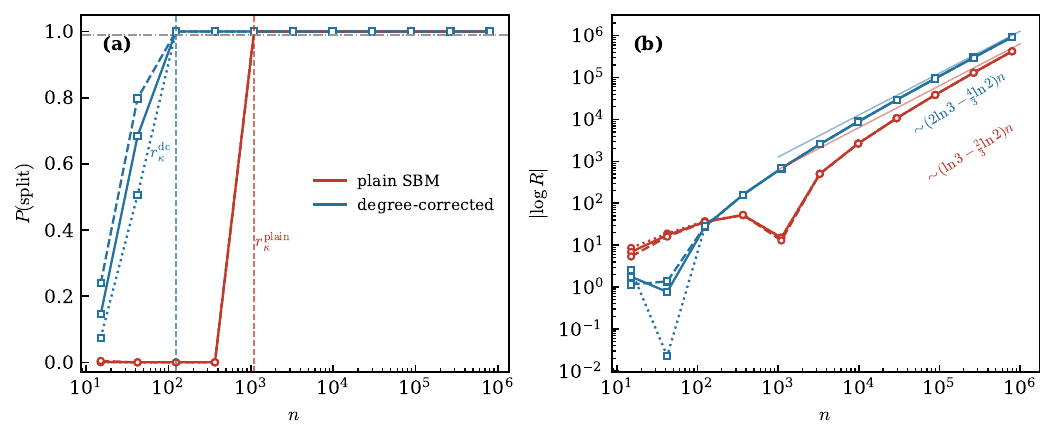}
\caption{(a) Posterior weight of the recursive branch partition,
Eq.~\eqref{eq:psplit}, versus network size $n$ (log scale). Red: plain
Bernoulli SBM; blue: degree-corrected (config-null) model. Line styles denote
$\alpha=0.5$ (dotted), $1$ (solid), $2$ (dashed). The dash-dotted line marks
$q=0.99$; vertical guides mark the Ramsey community numbers
$r_\kappa^{\mathrm{plain}}=1095$ and $r_\kappa^{\mathrm{dc}}=123$ at that
certainty. (b) The magnitude of the log-evidence ratio, $|\log R|$; both models
grow linearly in $n$, with prior-independent slopes
$\ln 3-\tfrac{2}{3}\ln 2$ (plain) and $2\ln 3-\tfrac{4}{3}\ln 2$
(degree-corrected), the latter exactly twice the former.}
\label{fig:main}
\end{figure*}

\section{Introduction}
\label{sec:intro}

Community structure---the organisation of nodes into groups more densely
connected internally than to the rest of the network---is one of the most
studied features of complex networks~\cite{girvan2002community,fortunato2010community,fortunato2016community}.
It is usually attributed to node heterogeneity, but node heterogeneity is not a
necessary condition: networks grown by purely local
rules~\cite{vazquez2003growing} routinely acquire communities as they
grow~\cite{vazquez2025ramsey}. To make this quantitative,
Ref.~\cite{vazquez2025ramsey} introduced the \emph{Ramsey community number}
$r_\kappa$, the minimum graph size that guarantees, with near-certainty, that a
prescribed detection method reports two or more communities; the name evokes
Ramsey theory, in which a large enough structure unavoidably contains ordered
substructures~\cite{ramsey1930problem}. A companion paper computed $r_\kappa$
analytically for the circulant ring lattice $C_n(1,\dots,c)$, a homogeneous
locally wired graph, using a Bernoulli stochastic block model (SBM) with
symmetric $\mathrm{Beta}$ priors as the detection rule: the plain cycle
($c=1$) is never partitioned ($r_\kappa=\infty$), while the next-nearest
neighbour ring ($c=2$) acquires a finite $r_\kappa$ of a few tens of nodes,
with a log-evidence growing as $(\ln 2)\,n$ \cite{vazquez2026ringwantsbroken}.

Those results rest on a special feature of the ring: it is
\emph{degree regular}. In a regular graph a two-block partition can only
capture assortative community structure, because there is no degree variation
to be explained away. Most networks of interest are not regular, and the
sharpest test of the emergence idea is a graph that is strongly heterogeneous,
where the naive question ``does it want to split?'' becomes confounded. On a
scale-free graph a plain block model will happily split the network into a
high-degree core and a low-degree periphery, not because the two groups form
communities but simply because they have different mean degrees. Distinguishing
genuine community structure from this degree artefact is the central issue of
degree-corrected community detection~\cite{karrer2011stochastic}.

Here I address both questions at once, analytically, for a canonical
deterministic scale-free graph: the pseudofractal web of Dorogovtsev, Goltsev,
and Mendes~\cite{dorogovtsev2002pseudofractal}. This graph is built by a fixed
recursive rule, has a power-law degree distribution with exponent
$\gamma=1+\ln 3/\ln 2\simeq 2.585$, and---crucially for an exact
treatment---has a natural recursive bipartition whose block sizes and edge
counts are known in closed form at every generation. I compute the Bayesian
evidence for this partition under two detection rules. The first is the plain
Bernoulli SBM. The second is a degree-corrected Poisson
SBM tested against the configuration-model null, which holds the degree
sequence fixed and asks whether the partition explains structure \emph{beyond}
what the degrees already imply. Both evidences are single closed-form
expressions, so the transition and $r_\kappa$ follow exactly.

The answers are clean and complementary. Under the plain SBM the web is
unpartitioned while small and breaks at $r_\kappa=1095$ nodes, with
$\log R\sim(\ln 3-\tfrac23\ln 2)n$. Under degree correction the same partition
breaks much earlier, $r_\kappa=42$, with a log-evidence growing exactly twice as
fast, $\log R_{\mathrm{dc}}\sim(2\ln 3-\tfrac43\ln 2)n$, and both slopes are
independent of the prior. Degree correction reverses the ordering that the
plain model assigns to the competing cuts: the hub--leaf split that the plain
model most prefers is demoted, and the recursive split is elevated. Pushing
this further, the self-similarity of the web makes the optimal description a
nested hierarchy of communities, refined down to $\sim\sqrt{n}$ blocks. The
recursion therefore builds true hierarchical community structure on top of its
scale-free degree sequence. The rest of the paper defines the web and the
partition (Sec.~\ref{sec:model}), derives the plain evidence ratio and its
transition (Sec.~\ref{sec:plain}), constructs the degree-corrected evidence and
shows the partition survives it (Sec.~\ref{sec:dc}), resolves the hierarchical
community structure (Sec.~\ref{sec:hier}), reports the Ramsey community numbers
(Sec.~\ref{sec:kappa}), and concludes (Sec.~\ref{sec:conclusions}).

\section{Model and detection rule}
\label{sec:model}

\subsection{The pseudofractal web}
\label{sec:web}

The pseudofractal scale-free web~\cite{dorogovtsev2002pseudofractal} is grown
deterministically. Start ($t=0$) from a triangle: three vertices and three
edges. At each subsequent generation, every existing edge spawns one new vertex,
joined to both endpoints of that edge; old edges are retained. Counting gives,
at generation $t$,
\begin{equation}
n_t=\frac{3}{2}\bigl(3^{t}+1\bigr),\qquad
E_t=3^{\,t+1},
\label{eq:NE}
\end{equation}
so the mean degree $2E_t/n_t\to 4$ and the graph is sparse. A vertex introduced
at generation $\tau$ has its degree doubled at every later generation, so
degrees take the values $2^{\,t-\tau+1}$ and their distribution is a power law
$P(k)\sim k^{-\gamma}$ with $\gamma=1+\ln 3/\ln 2\simeq 2.585$. The three seed
vertices are permanent hubs of maximal degree $2^{\,t+1}$.

\subsection{The recursive branch bipartition}
\label{sec:branch}

\begin{figure}[t]
\centering
\includegraphics[width=0.94\columnwidth]{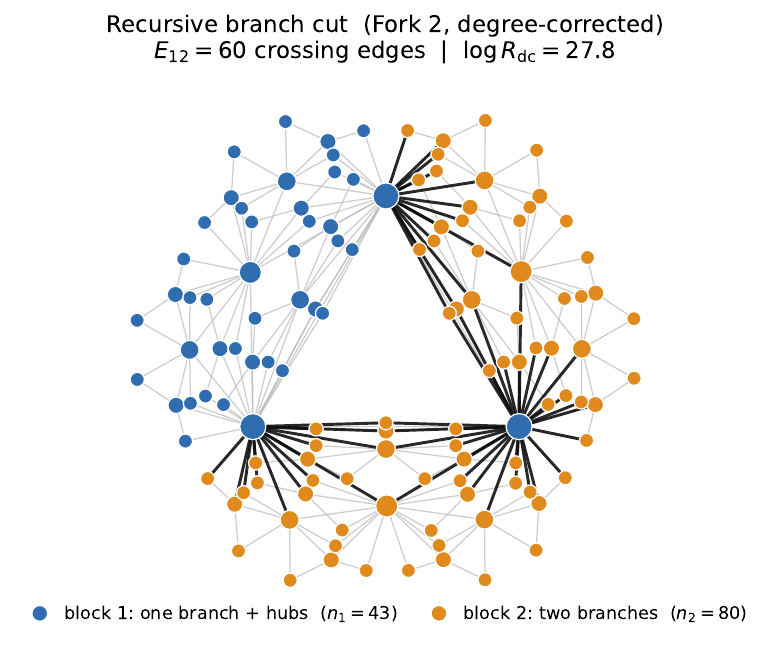}
\caption{The recursive branch partition of the pseudofractal web (generation
$t=4$, $n=123$; node area $\propto$ degree). One descent branch together with
the three seed hubs forms block~1; the other two branches form block~2. The two
blocks are joined by only $E_{12}=60$ crossing edges (highlighted), a fraction
$\sim(2/3)^{t}$ of all edges. This is the closed-form partition evaluated by
both detection rules, and the one the degree-corrected model favours.}
\label{fig:branch}
\end{figure}

Every non-seed vertex descends from exactly one edge, which in turn belongs to
one of the three ``branches'' rooted on the three sides of the initial triangle.
This assigns a branch label $b\in\{0,1,2\}$ to each of the $n_t-3$ non-seed
vertices, $(3^{t}-1)/2$ per branch, while the three seed hubs are shared. The
natural recursive bipartition places one branch together with the three hubs in
block~$1$, and the other two branches in block~$2$; it respects the
$\mathbb{Z}_2$ symmetry exchanging the two branches of block~$2$. 
The block sizes and the within/between edge counts are
all closed form (verified against direct construction through $t=8$):
\begin{align}
n_1&=\tfrac{1}{2}\!\left(3^{t}+5\right), & n_2&=3^{t}-1,\label{eq:sizes}\\
E_1&=3^{t}+2, & E_2&=2\cdot 3^{t}+2-2^{\,t+2},\label{eq:within}\\
E_{12}&=2^{\,t+2}-4, & &\label{eq:between}
\end{align}
with $n_1+n_2=n_t$ and $E_1+E_2+E_{12}=E_t=3^{t+1}$. The seam between the two
blocks is thin: $E_{12}=2^{t+2}-4$ grows only as $2^{t}=n^{\ln 2/\ln 3}\simeq
n^{0.631}$, so the fraction of crossing edges $E_{12}/E_t\sim(2/3)^{t}\to 0$.
The two branches are joined by a vanishing density of edges
(Fig.~\ref{fig:branch})---the structural fact that drives everything below.

\subsection{Stochastic block model evidence}
\label{sec:sbm}

The detection rule is Bayesian model comparison between the bipartition of
Sec.~\ref{sec:branch} and the unpartitioned network. Under a Bernoulli SBM with
within/within/between connection probabilities $\theta_1,\theta_2,\theta_{12}$
and a Bernoulli($\pi$) label prior, the marginal likelihood (evidence) of a
partition into blocks of sizes $(n_1,n_2)$ is
\begin{align}
P=\frac{1}{Z}\int
&\Bigl[\textstyle\prod_{i\in\{1,2\}}
\theta_i^{\,E_i}(1-\theta_i)^{\binom{n_i}{2}-E_i}\Bigr]
\nonumber\\
&\times\,\theta_{12}^{\,E_{12}}(1-\theta_{12})^{n_1 n_2-E_{12}}
\nonumber\\
&\times\,\pi^{n_1}(1-\pi)^{n_2}\;P(\theta)\,P(\pi)\,d\theta\,d\pi ,
\label{eq:model}
\end{align}
with symmetric priors $P(\theta_j)=\mathrm{Beta}(\theta_j;\alpha,\alpha)$
($j\in\{1,2,12\}$) and $P(\pi)=\mathrm{Beta}(\pi;\alpha,\alpha)$. Each
one-dimensional integral collapses to a Beta function through
$\int_0^1 x^{E+\alpha-1}(1-x)^{M-E+\alpha-1}dx=\Bt(E{+}\alpha,\,M{-}E{+}\alpha)$.
A block that carries no data integrates to a bare prior normaliser
$\Bt(\alpha,\alpha)$ that does not cancel, a Bayesian Occam penalty for unused
parameters. The unpartitioned network is the special case $n_1=0$: its two
empty blocks each leave one $\Bt(\alpha,\alpha)$, its single occupied block sees
all $E_t$ edges among $\binom{n_t}{2}$ pairs, and its label integral is
$\Bt(\alpha,n_t{+}\alpha)$.

\section{Plain evidence ratio and its transition}
\label{sec:plain}

The global constant $Z$ and the common prefactor cancel in the ratio
$R\equiv P_{\mathrm{split}}/P_{\mathrm{null}}$ of the branch partition to the
unpartitioned network,
\begin{widetext}
\begin{equation}
R=\frac{
\Bt\!\big(E_1{+}\alpha,\,\binom{n_1}{2}{-}E_1{+}\alpha\big)\,
\Bt\!\big(E_2{+}\alpha,\,\binom{n_2}{2}{-}E_2{+}\alpha\big)\,
\Bt\!\big(E_{12}{+}\alpha,\,n_1 n_2{-}E_{12}{+}\alpha\big)\,
\Bt\!\big(n_1{+}\alpha,\,n_2{+}\alpha\big)}
{\Bt(\alpha,\alpha)^{2}\,
\Bt\!\big(E_t{+}\alpha,\,\binom{n_t}{2}{-}E_t{+}\alpha\big)\,
\Bt\!\big(\alpha,\,n_t{+}\alpha\big)},
\label{eq:R}
\end{equation}
\end{widetext}
with the counts~\eqref{eq:sizes}--\eqref{eq:between} inserted. The posterior
weight of the partition, restricted to these two hypotheses, is
\begin{equation}
P(\mathrm{split})=\frac{R}{1+R}=\frac{1}{1+e^{-\log R}} .
\label{eq:psplit}
\end{equation}
The branch cut is
\emph{unbalanced} ($n_1\neq n_2$), so the label factor
$\Bt(n_1{+}\alpha,n_2{+}\alpha)$ carries a genuine cost; it is precisely this
term that supplies a transition. Evaluated in logarithmic form with
\SI{80}{digit} arithmetic, $\log R$ is negative and decreasing for small
generations and then turns sharply positive, Table~\ref{tab:plain}: the web is
\emph{not} partitioned while small, and flips to preferring the partition at
generation $t=6$, $n=1095$.

\begin{table}[t]
\centering
\begin{ruledtabular}
\begin{tabular}{c c d d d c}
$t$ & $n$ & \multicolumn{1}{c}{$\log R\ (\alpha{=}0.5)$}
& \multicolumn{1}{c}{$\log R\ (\alpha{=}1)$}
& \multicolumn{1}{c}{$\log R\ (\alpha{=}2)$}
& $P(\mathrm{split})$\\
\hline
3 & 42   & -19.01 & -17.33 & -16.01 & $\approx 0$\\
4 & 123  & -37.54 & -36.29 & -35.90 & $\approx 0$\\
5 & 366  & -52.72 & -52.08 & -52.91 & $\approx 0$\\
6 & 1095 &  15.20 &  15.17 &  12.98 & $\approx 1$\\
7 & 3282 & 506.25 & 505.52 & 501.90 & $\approx 1$\\
8 & 9843 &2667.04 &2665.58 &2660.51 & $\approx 1$\\
\end{tabular}
\end{ruledtabular}
\caption{Plain Bernoulli SBM: log-evidence ratio~\eqref{eq:R} for the recursive
branch partition of the pseudofractal web. The sign changes between $t=5$ and
$t=6$; $P(\mathrm{split})$ is for $\alpha=1$.}
\label{tab:plain}
\end{table}

The asymptotics are obtained from the counts~\eqref{eq:sizes}--\eqref{eq:between}
with Stirling's approximation. Writing $x\equiv 3^{t}$, every block is sparse
($E\sim x$ edges among $\sim x^2$ pairs), the seam $E_{12}\sim 2^{t}$ and the
null label are subleading, and the $x\log x$ and $x$ terms cancel between split
and null, leaving
\begin{equation}
\log R\;\sim\;\Bigl(\ln 3-\tfrac{2}{3}\ln 2\Bigr)\,n
\;\approx\;0.6365\,n ,
\label{eq:slope_plain}
\end{equation}
linear in $n$ and independent of the prior $\alpha$. The same scaling was obtained
for the partition of the ring \cite{vazquez2026ringwantsbroken}. A numerical fit of the discrete slope
$\Delta\log R/\Delta n$ converges to $0.636514$, matching
Eq.~\eqref{eq:slope_plain} to six digits, with the residual falling as
$(2/3)^{t}$.

\section{Degree-corrected evidence and the configuration-model null}
\label{sec:dc}

Because the web is scale free, the plain transition of Sec.~\ref{sec:plain} is
partly an artefact: a block model gains evidence simply by giving the
hub-rich and hub-poor regions different densities. To isolate genuine community
structure I replace the null by the configuration model, which fixes the degree
sequence, and the detection rule by a degree-corrected SBM.

\subsection{Construction}

Model the (sparse) graph as a Poisson multigraph with degree-corrected rates
$\langle A_{ij}\rangle=\theta_i\theta_j\,\omega_{g_i g_j}$ and
$\theta_i=k_i/\sqrt{2m}$, where $2m=\sum_i k_i=2E_t$ and $g_i$ is the block of
$i$~\cite{karrer2011stochastic}. With this choice $\omega_{rs}=1$ reproduces the
configuration model exactly, with expected block-pair edge counts
\begin{equation}
\Om_{rr}=\frac{\kappa_r^2}{4m},\qquad
\Om_{rs}=\frac{\kappa_r\kappa_s}{2m}\ (r\neq s),\qquad
\sum_{r\le s}\Om_{rs}=m,
\label{eq:omega}
\end{equation}
where $\kappa_r=\sum_{i\in r}k_i$ is the degree sum of block $r$. The Poisson
likelihood of the graph factorises as
\begin{equation}
P(A\mid\omega,g)=C(A,\bm{k})\prod_{r\le s}
\omega_{rs}^{\,e_{rs}}\,e^{-\omega_{rs}\Om_{rs}},
\label{eq:dclik}
\end{equation}
where $e_{rs}$ is the observed number of edges between blocks $r,s$, and the
prefactor $C(A,\bm{k})$ depends only on the (fixed) degrees and adjacency, not
on the partition $g$; it therefore \emph{cancels} in any evidence ratio between
partitions. Placing a symmetric conjugate prior
$\omega_{rs}\sim\mathrm{Gamma}(\alpha,\alpha)$ (mean $1$, centred on the
configuration model) and integrating,
\begin{align}
Z(e,\Om)&=\frac{\alpha^\alpha}{\Gamma(\alpha)}\int_0^\infty
\omega^{e+\alpha-1}e^{-(\Om+\alpha)\omega}\,d\omega
\nonumber\\
&=\frac{\alpha^\alpha}{\Gamma(\alpha)}\,
\frac{\Gamma(e{+}\alpha)}{(\Om{+}\alpha)^{e+\alpha}} .
\label{eq:Zdc}
\end{align}
The degree-corrected evidence ratio of the two-block partition to the
one-block configuration null is
\begin{equation}
R_{\mathrm{dc}}=\frac{Z(E_1,\Om_{11})\,Z(E_2,\Om_{22})\,Z(E_{12},\Om_{12})}
{Z(E_t,\Om_{\mathrm{tot}})},
\label{eq:Rdc}
\end{equation}
\begin{equation}
\Om_{\mathrm{tot}}=E_t,
\end{equation}
carrying an Occam factor $[\alpha^\alpha/\Gamma(\alpha)]^2$ for its two extra
affinities, exactly parallel to the $\Bt(\alpha,\alpha)^2$ penalty of the plain
model. For large arguments Eq.~\eqref{eq:Rdc} reduces, term by term, to the
degree-corrected assortativity signal
\begin{equation}
\log R_{\mathrm{dc}}\;\sim\;\sum_{r\le s} e_{rs}\,
\ln\!\frac{e_{rs}}{\Om_{rs}},
\label{eq:KL}
\end{equation}
the Kullback--Leibler surprise of the observed block-pair counts relative to
configuration-model expectations---the Bayesian, degree-corrected counterpart
of modularity~\cite{newman2006modularity}.

\subsection{The partition survives degree correction}

For the branch cut the block degree sums are again closed form,
\begin{equation}
\kappa_1=2\cdot 3^{t}+2^{\,t+2},\qquad
\kappa_2=4\cdot 3^{t}-2^{\,t+2},
\label{eq:kappas}
\end{equation}
with $\kappa_1+\kappa_2=2E_t$. Inserting~\eqref{eq:kappas} into
Eq.~\eqref{eq:omega} gives, in the large-$t$ limit, within-block densities that
\emph{exceed} the configuration-model expectation while the crossing density
vanishes:
\begin{equation}
\frac{E_1}{\Om_{11}}\to 3,\qquad
\frac{E_2}{\Om_{22}}\to \tfrac{3}{2},\qquad
\frac{E_{12}}{\Om_{12}}\sim\Bigl(\tfrac{2}{3}\Bigr)^{t}\!\to 0 .
\label{eq:ratios}
\end{equation}
Both branches hold more internal edges than their degrees alone would place,
and are joined by exponentially fewer crossing edges than expected: this is
assortative community structure that is not reducible to the degree sequence.
Consequently $\log R_{\mathrm{dc}}>0$ already from $t=3$ ($n=42$),
Table~\ref{tab:dc}, and grows as
\begin{equation}
\log R_{\mathrm{dc}}\;\sim\;\Bigl(2\ln 3-\tfrac{4}{3}\ln 2\Bigr)\,n
\;\approx\;1.2730\,n ,
\label{eq:slope_dc}
\end{equation}
again prior independent, and \emph{exactly twice} the plain
slope~\eqref{eq:slope_plain}, since
$2\ln 3-\tfrac43\ln 2=2(\ln 3-\tfrac23\ln 2)$. The leading term follows from
Eq.~\eqref{eq:KL} and the ratios~\eqref{eq:ratios}:
$E_1\ln 3+E_2\ln\tfrac32\to 3^{t}(3\ln 3-2\ln 2)
=(2\ln 3-\tfrac43\ln 2)\,n$. The numerical slope converges to $1.273028$, confirming
Eq.~\eqref{eq:slope_dc}.

\begin{table}[t]
\centering
\begin{ruledtabular}
\begin{tabular}{c c d d d}
$t$ & $n$ & \multicolumn{1}{c}{$\log R_{\mathrm{dc}}\ (\alpha{=}0.5)$}
& \multicolumn{1}{c}{$\log R_{\mathrm{dc}}\ (\alpha{=}1)$}
& \multicolumn{1}{c}{$\log R_{\mathrm{dc}}\ (\alpha{=}2)$}\\
\hline
2 & 15   &  -2.52 &  -1.77 &  -1.16\\
3 & 42   &   0.02 &   0.77 &   1.37\\
4 & 123  &  27.20 &  27.84 &  28.22\\
5 & 366  & 159.45 & 159.92 & 159.95\\
6 & 1095 & 682.44 & 682.70 & 682.32\\
7 & 3282 &2554.18 &2554.23 &2553.42\\
\end{tabular}
\end{ruledtabular}
\caption{Degree-corrected (configuration-null) log-evidence
ratio~\eqref{eq:Rdc} for the same branch partition. The sign changes between
$t=2$ and $t=3$: the partition is preferred at far smaller sizes than under
the plain model, Table~\ref{tab:plain}.}
\label{tab:dc}
\end{table}

\subsection{Degree correction reverses the preferred cut}

\begin{figure}[t]
\centering
\includegraphics[width=0.94\columnwidth]{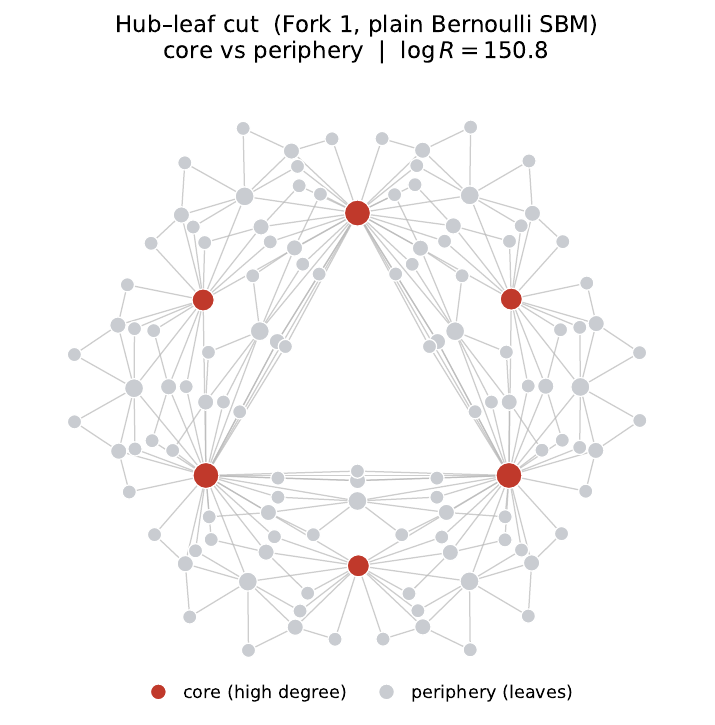}
\caption{The hub--leaf (core--periphery) partition of the same web
($t=4$), separating the highest-degree vertices (core) from the low-degree
leaves (periphery). The plain Bernoulli SBM rates this cut far above the
recursive one of Fig.~\ref{fig:branch}, but it merely tracks the degree
gradient: the configuration-model null absorbs it, and the degree-corrected
model demotes it (Table~\ref{tab:flip}).}
\label{fig:hubleaf}
\end{figure}

The two rules disagree about \emph{which} partition is best. Alongside the
recursive branch cut, consider the hub--leaf (core--periphery) cut that a plain
block model most prefers, obtained by thresholding on degree
(Fig.~\ref{fig:hubleaf}). Table~\ref{tab:flip} contrasts the two cuts under the
two nulls.

 The plain model rates the hub--leaf
cut far above the branch cut; degree correction inverts this, rating the
recursive branch cut well above the hub--leaf one. In other words, the large
plain evidence of the hub--leaf cut is mostly the degree gradient, which the
configuration null absorbs, whereas the branch cut encodes structure that
persists once degrees are fixed. A local search over all bipartitions under
$R_{\mathrm{dc}}$ finds an optimum that overlaps the branch cut by
$\approx 65\%$; the closed-form branch cut is thus a tractable witness of the
degree-corrected structure, not necessarily its global optimum.

\begin{table}[t]
\centering
\begin{ruledtabular}
\begin{tabular}{c c c d d}
$t$ & $n$ & cut & \multicolumn{1}{c}{plain $\log R$}
& \multicolumn{1}{c}{$\log R_{\mathrm{dc}}$}\\
\hline
5 & 366  & branch          & -52.08 & 159.92\\
5 & 366  & hub--leaf       & 508.74 &  19.32\\
7 & 3282 & branch          & 505.52 &2554.23\\
7 & 3282 & hub--leaf       &4844.88 & 209.25\\
\end{tabular}
\end{ruledtabular}
\caption{Which cut wins depends on the null ($\alpha=1$). The plain SBM prefers
the hub--leaf cut (a degree artefact); the degree-corrected model prefers the
recursive branch cut (genuine community structure).}
\label{tab:flip}
\end{table}

\section{Hierarchical community structure}
\label{sec:hier}

\begin{figure}[t]
\centering
\includegraphics[width=0.94\columnwidth]{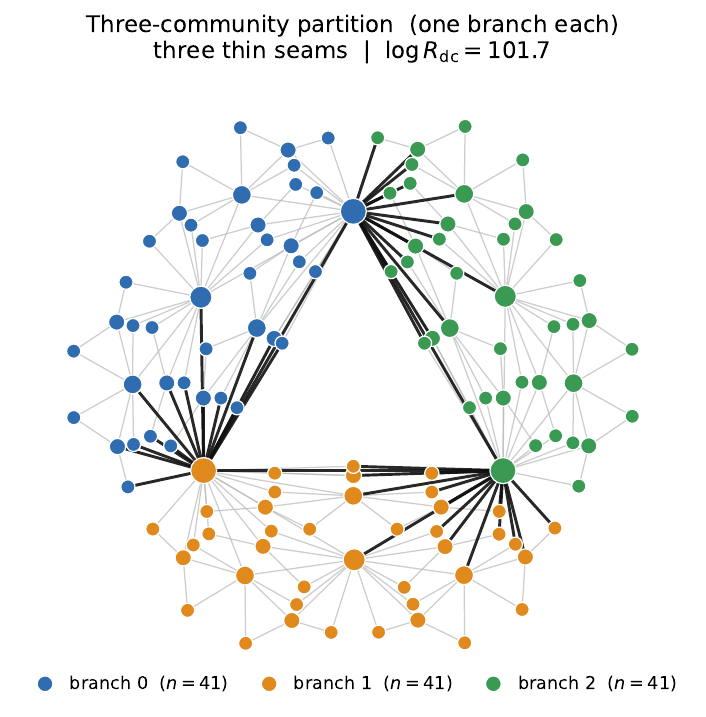}
\caption{The symmetric three-community partition of the web ($t=4$), one
descent branch per block. All three within-block densities exceed the
configuration-model expectation and all three seams (highlighted) vanish
as $(2/3)^{t}$, so the degree-corrected evidence,
Eq.~\eqref{eq:slope3}, beats the two-block cut of Fig.~\ref{fig:branch}.}
\label{fig:three}
\end{figure}

The branch cut of Sec.~\ref{sec:dc} lumps two of the three descent branches
into one block, so the thin seam between them is misread as within-block
structure. Separating all three branches should therefore fit better. Placing
each branch with one seed hub in its own block gives a symmetric
three-community partition whose counts are again closed form (verified through
$t=8$):
\begin{equation}
n_r=\tfrac{1}{2}(3^{t}{+}1),\quad \kappa_r=2\cdot 3^{t},\quad
e_{rr}=3^{t}{-}2^{t},\quad e_{rs}=2^{t},
\label{eq:three_counts}
\end{equation}
for each block $r$ and each pair $r\neq s$. With
$\Omega_{rr}=3^{t}/3$ and $\Omega_{rs}=2\cdot 3^{t}/3$ from
Eq.~\eqref{eq:omega}, every block is assortative and every seam vanishing,
\begin{equation}
\frac{e_{rr}}{\Omega_{rr}}=3\bigl(1-(2/3)^{t}\bigr)\to 3,\qquad
\frac{e_{rs}}{\Omega_{rs}}=\tfrac{3}{2}(2/3)^{t}\to 0 ,
\label{eq:three_ratios}
\end{equation}
a genuine three-community structure. Its degree-corrected evidence exceeds the
two-block value at every size (Fig.~\ref{fig:three}) and grows as
\begin{equation}
\log R_{\mathrm{dc}}^{(3)}\sim 2\ln 3\,n\approx 2.197\,n ,
\label{eq:slope3}
\end{equation}
larger than the two-block slope~\eqref{eq:slope_dc} by exactly
$\tfrac{4}{3}\ln 2\,n$---the evidence the branch cut forfeited by merging two
communities.

The construction is self-similar, so the refinement does not stop at three.
Each branch is itself a pseudofractal web grown from an edge, and removing the
generation-$d$ edge set partitions the graph into $3^{\,d+1}$ statistically
identical sub-webs. The degree-corrected evidence keeps rising as the
resolution is refined (Table~\ref{tab:hier}); at fixed level $d$ it grows as
\begin{equation}
\log R_{\mathrm{dc}}^{(d)}\sim 2(d{+}1)\ln 3\,n ,
\label{eq:sloped}
\end{equation}
so each additional level adds a further $\simeq 2\ln 3\,n$. Deeper partitions
win until the blocks shrink to where the seams and the Occam penalty are no
longer negligible. Over the accessible range $T\le 9$ the optimum sits at
\begin{equation}
K_{\mathrm{opt}}=3^{\lfloor t/2\rfloor}\sim\sqrt{n},
\label{eq:kopt}
\end{equation}
so the web is best described by of order $\sqrt{n}$ communities of $\sim\sqrt{n}$
nodes each---the geometric midpoint of its own hierarchy. The single branch cut
of Sec.~\ref{sec:dc} is merely the top of this tree; the full description is the
nested, hierarchical block structure characteristic of self-similar
graphs~\cite{peixoto2014hierarchical}.

\begin{table}[t]
\centering
\begin{ruledtabular}
\begin{tabular}{c c r r r r c}
$t$ & $n$ & \multicolumn{1}{c}{$K{=}3$} & \multicolumn{1}{c}{$K{=}9$}
& \multicolumn{1}{c}{$K{=}27$} & \multicolumn{1}{c}{$K{=}81$}
& $K_{\mathrm{opt}}$\\
\hline
4 & 123   & 102   & 164    & 88     & $-37$  & 9\\
5 & 366   & 436   & 826    & 796    & 360    & 9\\
6 & 1095  & 1601  & 3190   & 3970   & 2939   & 27\\
7 & 3282  & 5458  & 11013  & 15334  & 15042  & 27\\
8 & 9843  & 17823 & 36005  & 52453  & 60558  & 81\\
9 & 29526 & 56658 & 114244 & 169391 & 211877 & 81\\
\end{tabular}
\end{ruledtabular}
\caption{Degree-corrected log-evidence $\log R_{\mathrm{dc}}$ ($\alpha=1$) for
the hierarchical partition into $K=3^{\,d+1}$ communities ($K{=}3$: the
symmetric partition of Eq.~\eqref{eq:three_counts}; deeper levels: each
generation-$d$ sub-web, with shared vertices attached to the majority
neighbouring block). The maximum in each
row (column $K_{\mathrm{opt}}$) shifts to finer resolution as the web grows,
tracking $K_{\mathrm{opt}}=3^{\lfloor t/2\rfloor}\sim\sqrt{n}$,
Eq.~\eqref{eq:kopt}.}
\label{tab:hier}
\end{table}

\section{The Ramsey community number}
\label{sec:kappa}

Following~\cite{vazquez2025ramsey}, the Ramsey community number is the minimum
size at which the connectivity is preferentially described by communities to a
prescribed certainty $q$,
\begin{equation}
r_\kappa(q)=\min\bigl\{\,n_t:\ P(\mathrm{split})\ge q\,\bigr\},
\end{equation}
with $P(\mathrm{split})$ from the relevant evidence ratio. Because the web
exists only at the discrete sizes $n_t$ of Eq.~\eqref{eq:NE}, $r_\kappa$ takes
one of those values. The results are collected in Table~\ref{tab:kappa}. Under
the plain model the transition is so abrupt---$\log R$ leaps from $-52$ at
$t=5$ ($n=366$) to $+15$ at $t=6$ ($n=1095$) to $+506$ at $t=7$ ($n=3282$)---that $r_\kappa=1095$ for every
certainty level up to $q\simeq 1-10^{-5}$ and every prior $\alpha$. Under degree correction the web
breaks at $r_\kappa=42$ ($q=\tfrac12$) or $123$ ($q\ge 0.9$), again essentially
$\alpha$ independent.
\begin{equation}
r_\kappa^{\mathrm{plain}}=1095,\qquad
r_\kappa^{\mathrm{dc}}=42\text{--}123 .
\end{equation}

\begin{table}[t]
\centering
\begin{ruledtabular}
\begin{tabular}{c c c c c c}
model & $\alpha$ & $r_\kappa^{0.5}$ & $r_\kappa^{0.9}$
& $r_\kappa^{0.99}$ & $r_\kappa^{0.999}$\\
\hline
plain & 0.5 & 1095 & 1095 & 1095 & 1095\\
plain & 1.0 & 1095 & 1095 & 1095 & 1095\\
plain & 2.0 & 1095 & 1095 & 1095 & 1095\\
\hline
dc & 0.5 & 42 & 123 & 123 & 123\\
dc & 1.0 & 42 & 123 & 123 & 123\\
dc & 2.0 & 42 & 123 & 123 & 123\\
\end{tabular}
\end{ruledtabular}
\caption{Ramsey community number $r_\kappa^{q}$ (smallest generation size $n_t$
with $P(\mathrm{split})\ge q$) for the pseudofractal web, under the plain
Bernoulli SBM and the degree-corrected (dc) configuration-null model. Degree
correction lowers $r_\kappa$ by more than an order of magnitude and is nearly
prior independent.}
\label{tab:kappa}
\end{table}

The two numbers make the interpretation explicit. The plain
$r_\kappa=1095$ conflates two effects and switches on only when the accumulated
degree-plus-community signal wins; the degree-corrected $r_\kappa=42$ measures
the community signal alone, which is present almost from the start. That the
degree-corrected threshold is the \emph{smaller} of the two shows the recursion
imprints community structure early and the degree gradient, if anything, delays
the plain detector rather than driving it.

\section{Conclusions}
\label{sec:conclusions}

I have computed the Ramsey community number of the deterministic pseudofractal
scale-free web analytically, under two detection rules whose Bayesian evidences
are both closed form: a plain Bernoulli SBM and a degree-corrected Poisson SBM
tested against the configuration-model null. The natural recursive bipartition
of the web has block sizes and edge counts known exactly at every generation,
Eqs.~\eqref{eq:sizes}--\eqref{eq:between} and \eqref{eq:kappas}, which makes
both evidence ratios explicit.

The results are fivefold. (1)~Under the plain SBM the web is unpartitioned while
small and acquires a finite Ramsey community number $r_\kappa=1095$, above which
$\log R\sim(\ln 3-\tfrac23\ln 2)n$, prior independent. (2)~Under degree
correction the same partition survives---its within-block densities exceed, and
its crossing density falls exponentially below, configuration-model
expectations, Eq.~\eqref{eq:ratios}---and breaks at the much smaller
$r_\kappa=42$, with $\log R_{\mathrm{dc}}\sim(2\ln 3-\tfrac43\ln 2)n$, exactly
twice the plain slope. (3)~Degree correction reverses the ordering of the
candidate cuts, demoting the hub--leaf split favoured by the plain model and
elevating the recursive one, so the effect is genuine community structure, not
degree heterogeneity. (4)~Both transitions are prior independent in their
asymptotic slope, and the discrete generation sizes make $r_\kappa$ a sharp step.
(5)~Because the web is self-similar, the degree-corrected evidence keeps rising
as the partition is refined---each hierarchical level adding $\simeq2\ln 3\,n$,
Eq.~\eqref{eq:sloped}---so the optimal description is not two or three
communities but a nested hierarchy of
$K_{\mathrm{opt}}=3^{\lfloor t/2\rfloor}\sim\sqrt{n}$ blocks.

Together with the companion ring calculation \cite{vazquez2026ringwantsbroken}, these results extend the exactly
solvable theory of community emergence from a homogeneous graph to a strongly
heterogeneous one, and supply the degree-correction control that a scale-free
setting demands. They confirm that a purely local recursive rule can build
hierarchical communities over and above the degree sequence it also produces.
Natural extensions include a proof that $K_{\mathrm{opt}}\sim\sqrt{n}$
asymptotically together with the exact evidence at the optimal resolution, other
deterministic fractals such as the $(u,v)$-flowers, and the scaling of
$r_\kappa$ with the branching number---each within reach of the present
closed-form approach.

\begin{acknowledgments}
The calculations, computer scripts, and text of this manuscript were generated
by Claude Opus 4.8 (Anthropic). The analytical results were verified with an
accompanying open-source script at 
\href{https://github.com/av2atgh/pseudofractalweb}{github.com/av2atgh/pseudofractalweb},
and the closed-form edge and degree counts
were checked against direct construction of the web through eight generations.
\end{acknowledgments}

\bibliographystyle{apsrev4-2}
%

\end{document}